\documentclass[pra,aps,superscriptaddress,twocolumn]{revtex4}
\usepackage{latexsym}
\usepackage{amssymb}
\usepackage{amsmath}
\usepackage{graphicx}
\usepackage{dcolumn}
\usepackage{bm}
\usepackage{amsfonts}
\usepackage{color}
\usepackage{epsfig}
\usepackage{color}

\begin{document}
\title{Quantum phases of spinful Fermi gases in optical cavities}

\author{E. Colella}
\affiliation{Dipartimento di Fisica ``Enrico Fermi'', Universit\`a di Pisa and INFN,
  Largo B. Pontecorvo 3, I-56127 Pisa, Italy}
\affiliation{Institut f\"{u}r Theoretische Physik, Universit\"{a}t Innsbruck,
  Technikerstrasse 21/3, A-6020 Innsbruck, Austria}

\author{R. Citro}
\affiliation{Dipartimento di Fisica ``E.R. Caianiello'', Universit\`a
  degli Studi di Salerno, Via Ponte Don Melillo, I-84084 Fisciano (Sa), Italy}
\affiliation{INFN-Sezione di Pisa, Largo B. Pontecorvo 3, I-56127 Pisa, Italy}

\author{M. Barsanti}
\affiliation{Dipartimento di Ingegneria Civile e Industriale, Universit\`a di Pisa and INFN,
  Largo L. Lazzarino, I-56122 Pisa, Italy}

\author{D. Rossini}
\affiliation{Dipartimento di Fisica ``Enrico Fermi'',
  Universit\`a di Pisa and INFN, Largo B. Pontecorvo 3, I-56127 Pisa, Italy}

\author{M.-L. Chiofalo}
\affiliation{Dipartimento di Fisica ``Enrico Fermi'', Universit\`a di Pisa and INFN,
  Largo B. Pontecorvo 3, I-56127 Pisa, Italy}

\begin{abstract}
  We explore the quantum phases emerging from the interplay between spin and motional degrees of freedom of a
  one-dimensional quantum fluid of spinful fermionic atoms, effectively interacting via a photon-mediating mechanism with tunable sign and strength $g$,
  as it can be realized in present-day experiments with optical cavities. We find the emergence, in the very same system,
  of spin- and atomic-density wave ordering, accompanied by the occurrence of superfluidity for $g>0$, while cavity photons are seen
  to drive strong correlations at all $g$ values, with fermionic character for $g>0$, and bosonic character for $g<0$. Due to the long-range nature
  of interactions, to infer these results we combine mean-field and exact diagonalization methods supported by bosonization analysis.
\end{abstract}

\date{\today}

\maketitle

\section{Introduction}
Quantum-degenerate atomic gases represent paradigmatic systems
for many-body and fundamental physics applications, as extreme quantum conditions
can be engineered with remarkably high accuracy and precision~\cite{RMPGiorginiStringari, Dalfovo, Dalibard}.
In this context, a particularly versatile platform is provided by atoms in optical cavities,
where light-matter coupling can be greatly enhanced~\cite{EsslingerReview, Vuletic2011}.
Long-range mediated interactions among the atoms affect both atomic internal and external degrees of freedom
leading to superradiance, a self-organization of the atomic density above a critical
pump strength~\cite{DomokosRitsch2002, Nagy2008, KBS2014, Sandner2015, maschlermekhov, MaschlerRitsch2005}.

The transition to a superradiant state was experimentally observed both in thermal gases and in
BECs~\cite{Wolke,Brenneke2007,Black2003,Baumann2010,Baumann2011,Esslinger12}. For fermionic atoms,
a striking suppression of the critical pump strength was predicted, when the cavity-photon recoil equals
$2k_F$~\cite{Piazza}, and topological phase transitions for spinless fermions were studied~\cite{Farokh2016}.
The possibility of realizing systems with two effective spin states upon choosing suited transitions between
internal atomic states opens the possibility to explore the physics of pseudospins coupled to
a cavity mode~\cite{Kimble}. Topological phase transitions were studied for spinful gases
too~\cite{Pan2015, Yu2018, Farokh2}, leading to superradiance and spin waves.
Interactions between cavity pseudospins were proposed as a mechanism for realizing
BCS-BEC crossover~\cite{Guo2012} and generating spin-orbit coupling~\cite{Deng2014, Yu2018}.

So far, the possible competition between density and spin-driven processes in shaping the quantum phase diagram,
along with the peculiar strong correlations effects manifesting in reduced dimensions,
have been largely unexplored.
Here we combine mean-field (MF) and beyond MF methods, i.e., exact diagonalization (ED)
supported by bosonization analysis, to unveil such novel quantum effects, focusing on a one-dimensional (1D) geometry.
These are driven by the interplay between internal and external degrees of freedom, and quantum fluctuations
originated by reduced dimensionality and tunable strengths of the effective atomic interaction $g$.
Cavity photons are able to mediate strong correlations at all, even tiny, coupling strengths.
From ED and bosonization analysis of the fluid structure,
we find the emergence of spin-density waves (SDW) for positive and negative $g$ values,
accompanied by atomic-density wave (ADW) ordering for $g>0$. The quantum fluid displays boson-like character for $g<0$,
and fermion-like for $g>0$. In the latter region, bosonization and renormalization group
suggest that a superfluid (SF) state may occur. While the long-range nature of photon-mediated interactions
prevents us from using more refined numerical methods as the density-matrix renormalization group (DMRG),
a combined use of ED and bosonization allows us to cross-check the structure of the fluid and to explore different
aspects of the underlying physics beyond MF.

The paper is organized as follows. In Sec.~\ref{sec:model} we set in the system concept and derive the effective many-fermion Hamiltonian which embodies the
cavity-photon mediated interactions, which will be then investigated by various methods. In particular, Sec.~\ref{sec:MF} provides a quantitative flavor of the
rich phase diagram, as it can be worked out according to a MF treatment able to include spin-like and superfluid-like order parameters. In Sec.~\ref{sec:ED}, the
boson and fermion-like characteristics of the fluid structure emerging from the MF analysis are confirmed by means of ED of the effective
Hamiltonian, along with the SDW nature of the excitations and their primary role in originating any other indirect effect. In fact, the ED
analysis provides as well information on the expected existence of ADW excitations. Due to the use of canonical ensemble for memory-size practical
reasons, ED cannot provide reliable clues on the appearance of SF, as predicted by MF analysis. The final part of Sec.~\ref{sec:ED}
is devoted to a preliminary discussion of the results of a bosonization analysis for the same effective Hamiltonian, which confirm the ED
outcomes on the structure of the fluid,
and support the emergence of a SF in the parameter region inferred by the MF treatment. Finally, Sec.~\ref{sec:conclusions} summarizes the main
results of this work, their possible implementation and observation in current experiments with ultracold atoms in optical cavities,
and perspectives aimed at improving the
present investigation to include the intrinsic open nature of the cavity system, and to exploit our findings in applications.

\section{Model}\label{sec:model}
\begin{figure}[htb]
  \includegraphics[width=\columnwidth]{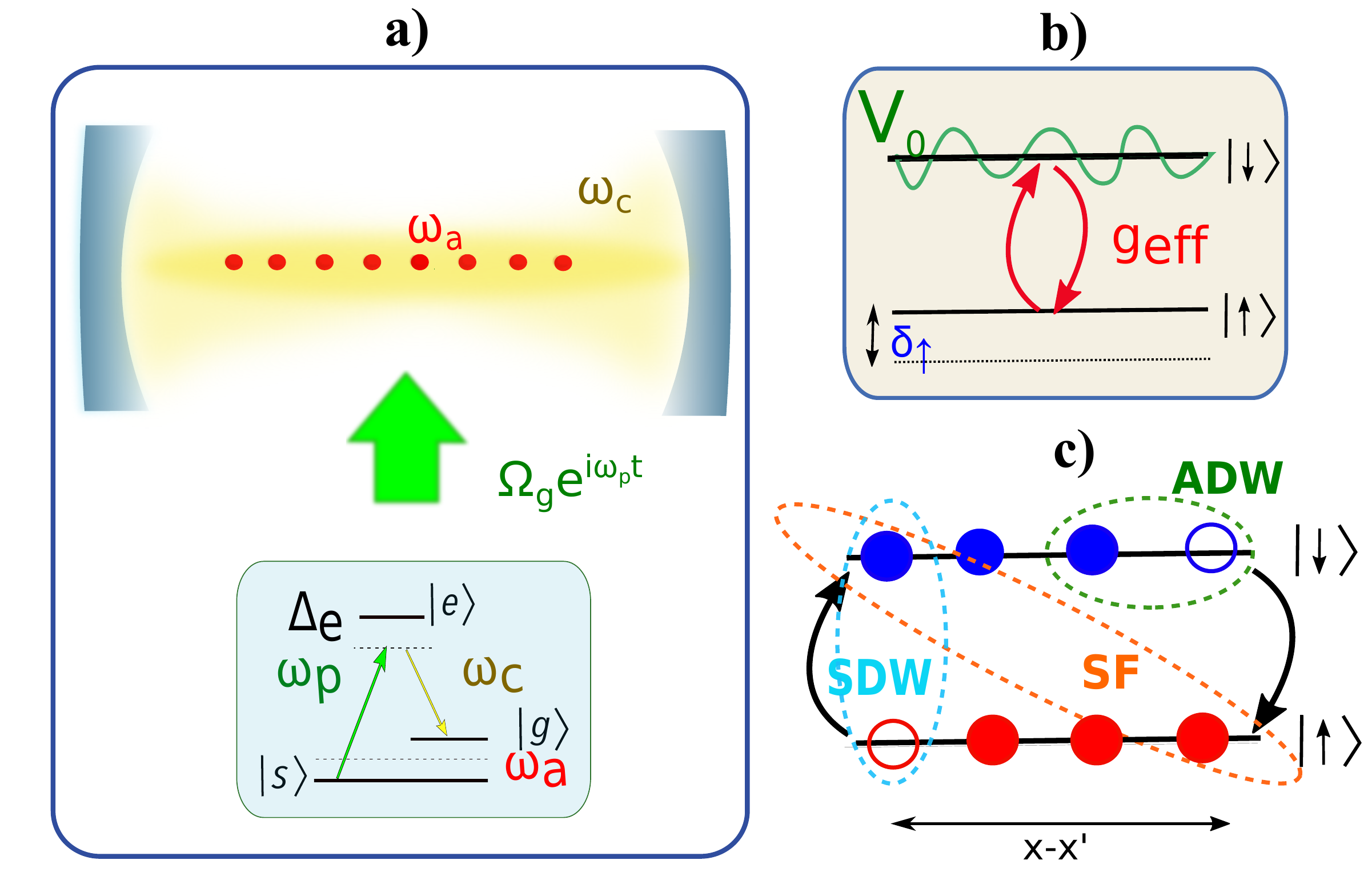}
  \caption{{\bf a)} System concept. Atoms in a $\Lambda$-scheme involving three levels
      ($|{s}\rangle$, $|{g}\rangle$, $|{e}\rangle$) are placed into a linear optical cavity
      with vacuum-mode frequency $\omega_c$, and transversely pumped by a classical field with frequency $\omega_p$
      and strength $\Omega$.
      {\bf b)} A large detuning $\Delta_e$ allows the adiabatic elimination of $|e\rangle$; the remaining
      effective levels $|{s}\rangle \equiv |\!\!\uparrow\rangle$ and $|{g}\rangle \equiv |\!\!\downarrow \rangle$,
      separated by a transition frequency $\omega_a$, are coupled to the cavity via $\hbar g_{\rm eff}$.
      The shift $\delta_{\uparrow}$ and potential $V_0(x)$ are embodied in the effective system.
      {\bf c)} Cartoon of Spin-Density-Wave (SDW), Atomic-Density-Wave (ADW), and Superfluid (SF)
      orderings expected from the resulting Hamiltonian.}
  \label{fig:fig1}
\end{figure}

Our system concept is sketched in Fig.~\ref{fig:fig1}.
An ensemble of three-level atoms in the so-called $\Lambda$-scheme is placed in a linear cavity characterized by
a vacuum mode of frequency $\omega_c$ and cavity loss $\kappa$. The atomic transition is transversely pumped by a classical field
with frequency $\omega_p$, connecting a state $|{s}\rangle$ to an excited state $|{e}\rangle$ with energy $\hbar \omega_e$.
The cavity field induces a transition between $|{e}\rangle$ and a third energy level, say
$|g\rangle$. After a unitary transformation of the resulting Hamiltonian $\hat H_\Lambda$
in the co-rotating frame of the pump, and adiabatically eliminating the state $|e\rangle$
under conditions of large detuning $\Delta_e \equiv \omega_p-\omega_e$,
the effective system turns out to be as in Fig.~\ref{fig:fig1}b): two-level atoms, labeled by
$|\!\!\uparrow\rangle$ and $|\!\!\downarrow\rangle$, interact with the cavity via an effective
two-photon Rabi coupling $\hbar g_{\rm eff} \equiv \hbar g_0 \Omega / \Delta_e$,
with $\hbar g_0$ being the original cavity-mode strength.
To enlighten the interplay between spin and motional degrees of freedom, we operate a few simplifications.
We neglect the classical shift $\delta_\uparrow=\hbar\Omega^2/\Delta_e$ induced by the two-photon
transition on the effective ground-state, which could be compensated by inducing a second light shift on $|\!\!\downarrow\rangle$.
We take the periodic potential experienced by the upper level to be small
and include its effect in an overall shift of the cavity resonance frequency. Finally,  we tune $\omega_p=\omega_a$ at resonance with the atomic transition.

We then consider conditions in which the photon dynamics, evolving on the time-scale
$\sim |\kappa+i(\omega_p-\omega_c)|^{-1}$, is much faster than the atomic dynamics,
evolving as $\sim \big| {\rm max} (g_{\rm eff}\sqrt{N},\Gamma) \big|^{-1}$ in terms of the number of atoms $N$
and the atomic transition width $\Gamma$. Therefore, the cavity field can be adiabatically eliminated by standard
procedures~\cite{GardinerZoller}. As we are interested in the emergence of strong correlations,
we focus on a 1D geometry, where a tight confinement is operated along the directions
perpendicular to the cavity axis. 
This can be achieved by means of a superimposed trapping due to either harmonic or two-dimensional
optical lattice. Provided that the recoil induced by the transverse pump is negligible, the transversal degrees of freedom can be integrated out,
and the system may exhibit a quasi-1D behavior. 
The effective many-body Hamiltonian for interacting fermions is:
\begin{eqnarray}
  \label{eq:H}
  \hat H & = & \sum_{\sigma=\uparrow,\downarrow}\int dx \, \hat \Psi^\dagger_{\sigma}(x) \left[-\frac{\hbar^2\nabla^2}{2m}\right] \hat \Psi_{\sigma}(x)\\
  &&+
  \hbar \, g\int dx \, dx' \, V(x,x') \hat \Psi^\dagger_\downarrow(x)\hat \Psi_\uparrow(x)\hat \Psi^\dagger_\uparrow(x')\hat \Psi_\downarrow(x'),\nonumber
\end{eqnarray}
with $\hat \Psi_\sigma^{(\dagger)}(x)$ satisfying usual anticommutation relations, $V(x,x') = \cos(k_Lx)\cos(k_Lx')$,
and $k_L$ being the laser wave vector.
The coupling strength $g = g_{\rm eff}^2\,{\Delta_c}/({\kappa^2/4+\Delta_c^2})$
can be experimentally varied in sign and strength by only tuning the parameter $\Delta_c$.

A glance at the interaction term in Eq.~\eqref{eq:H} unveils the relevant processes we expect to occur,
driven by spin $\big[ \sim \hat \Psi^\dagger_\downarrow(x)\hat \Psi_\uparrow(x) \big]$ and number density
$\big[ \sim \hat \Psi^\dagger_\sigma(x)\hat \Psi_\sigma(x') \big]$ fluctuations,
and of possible superfluid $\big[ \sim \hat \Psi^\dagger_\downarrow(x)\hat \Psi^\dagger_\uparrow(x') \big]$ pairing. In fact,
as represented by the cartoon in Fig.~\ref{fig:fig1}c), the investigation of the relationship among SDW, ADW,
and SF-like fluctuations in shaping the phase diagram of the quantum fluid, represents the main content of this work.
To provide a flavor of the underlying physics, we proceed by steps.
We first explore the interplay of SDW and SF-like ordering
within a MF approach, where the ADW ordering is temporarily frozen: considering the long-range nature
of the effective interactions, one may expect the MF approach able to provide the main picture. Then, we dig in
using ED and bosonization, to explore qualitative and quantitative changes to the emerging picture beyond MF.

\section{Mean-field treatment}\label{sec:MF}
Hamiltonian~\eqref{eq:H} can be simplified in MF approximation.
Since we are interested in the behavior of low-momentum pairs, we retain only
momentum-conserving terms.
Apart from a constant contribution to the ground state, in $k$-space and in the grand-canonical ensemble
one obtains~\cite{GardinerZoller, Guo2012, refnoteHelmut}:
\begin{eqnarray}
\hat H_{\rm MF} & = & \sum_{k, \sigma} \varepsilon_k \, \hat c_{k,\sigma}^\dagger \hat c_{k,\sigma} 
   - \!\!\! \sum_{k, \, Q= \pm k_L} \!\! S(Q) \, \hat c_{k,\downarrow}^{\dagger} \hat c_{k+Q,\uparrow}^{} \nonumber \\
&& +\sum_{k}\Delta(k)\big( \hat c_{-k,\downarrow}\hat c_{k,\uparrow}+ {\rm H.c.} \big) .
  \label{eq:Ham_MF}
\end{eqnarray}
Here, $\varepsilon_{k}\equiv \hbar^2 k^2/(2m)-\mu$ denotes the single particle kinetic energy referred
to the chemical potential $\mu$, determined after imposing conservation of the number of atoms,
while $S(q)$ and $\Delta(k)$ are the relevant order parameters according to the standard MF prescription.
Namely, the SF order parameter is defined as $\Delta(k)=-g\sum_{k'}V_{k, k'} \langle \hat c_{- k',\downarrow}\hat c_{k',\uparrow}\rangle$.
$V_{k, k'}=\delta(k'- k+k_L)+\delta(k'- k - k_L )$ is the microscopic interaction potential of Eq.~\eqref{eq:H},
representing the pairing of fermions with opposite spins and momenta eventually leading to macroscopic
ground-state occupation. Conversely, the SDW order parameter
$S(Q)=-g\sum_k \langle \hat c_{k,\downarrow}^{\dagger}\hat c_{k+Q,\uparrow}^{}\rangle$
describes spin waves propagating with wave-vector $Q=\pm k_c$,
made available by exchange of a cavity photon.
For simplicity we have taken both order parameters to be real
and disregarded Hartree-Fock terms.

Let us now concentrate on the ground-state properties of Hamiltonian~\eqref{eq:Ham_MF}.
By means of Green's functions techniques in imaginary time $\tau$, we can derive a set of self-consistent
equations for $\Delta(k)$, $S(Q)$, and $N$ in the 1D geometry under consideration. To this aim, the equations of motion
for the single creation and annihilation operators are derived and used to obtain the Green's functions evolution in imaginary time.
Unlike the standard BCS case though, here we have four Green's functions:
\begin{eqnarray}
\mathcal{G}(p, \tau-\tau') & = & - \big\langle \vec T_\tau \big[ \hat c_{p,\sigma}(\tau)\, \hat c^\dagger_{p,\sigma}(\tau') \big] \big\rangle, \\
  \mathcal{F}(p, \tau-\tau') & = & -\big\langle \vec T_\tau \big[ \hat c_{p,\uparrow}(\tau) \, \hat c_{-p,\downarrow}(\tau') \big] \big\rangle, \\
  \mathcal{S}_Q (p, \tau-\tau') & = & - \big\langle \vec T_\tau \big[ \hat c_{p,\uparrow}(\tau) \, \hat c^\dagger_{p+Q,\downarrow}(\tau') \big] \big\rangle, \\
  \mathcal{F}_{\uparrow\uparrow,Q}(p, \tau-\tau') & = & - \big\langle \vec T_\tau \big[ \hat c_{k,\uparrow}(\tau) \, \hat c_{-k-Q,\uparrow}(\tau') \big] \big\rangle,
\end{eqnarray}
with $Q=\pm k_L$, leading to six coupled equations.
In particular, keeping $\mathcal{F}_{\uparrow\uparrow,Q}$ ($\mathcal{F}_{\downarrow\downarrow,Q}$) is of pivotal importance to establish
self-consistency between spin and superfluid order parameters. The resulting coupled equations of motion are:
\begin{widetext}
\begin{eqnarray}
  \hbar \frac{\partial}{\partial \tau} \mathcal{G}(p,\tau-\tau') & = &
  -\hbar \, \delta(\tau-\tau')-\epsilon_{p}\, \mathcal{G}(p,\tau-\tau') +\sum_{q}S(-q) \, \mathcal{S}^\dagger_q(p,\tau-\tau')-\Delta(p) \, \mathcal{F}^\dagger(p,\tau-\tau'), \\
  \hbar \frac{\partial}{\partial \tau} \mathcal{F}^\dagger(p,\tau-\tau') & = &
  \epsilon_p \, \mathcal{F}^\dagger(p,\tau-\tau') -\sum_{q=\pm k_L}S(-q) \, \mathcal{F}^\dagger_{\uparrow\uparrow, q}(p,\tau-\tau')-\Delta^\dagger(p) \, \mathcal{G}(p,\tau-\tau'), \\
  \hbar \frac{\partial}{\partial \tau} \mathcal{S}_Q^\dagger(p,\tau-\tau') & = &
  -\epsilon_{p+Q} \, \mathcal{S}_Q^\dagger(p,\tau-\tau') +S^\dagger(-Q) \, \mathcal{G}(p,\tau-\tau')+\Delta(p+Q) \, \mathcal{F}^\dagger_{\uparrow\uparrow, Q}(p,\tau-\tau'), \\
  \hbar \frac{\partial}{\partial \tau} \mathcal{F}^\dagger_{\uparrow\uparrow, Q}(p,\tau-\tau') & = &
  \epsilon_{p+Q} \, \mathcal{F}^\dagger_{\uparrow\uparrow, Q}(p,\tau-\tau') -S^\dagger(-Q) \, \mathcal{F}^\dagger(p,\tau-\tau')+\Delta^\dagger(p+Q)\,\mathcal{S}^\dagger_Q(p,\tau-\tau').
\end{eqnarray}
This set of coupled equations can be cast into an algebraic linear system where the unknown variables are exactly the correlation functions.
The equations are solved in the frequency space rather than in the time domain, introducing the Fourier transform of each correlation function.
The resulting linear system is of the form $\mathbf{AX}=\mathbf{B}$, with
\begin{equation*}
  \mathbf{A}=
  \begin{pmatrix}
    \begin{tabular}{  c c c c c c }
      $i\omega_n-\epsilon_p$ & $S(-k_c)$ & $S(\small{k_c})$ & 0 & 0 &$-\Delta(p)$ \\
      $S(-k_c)$ & $i\omega_n-\epsilon_{p+k_c}$ & 0 & 0 & $\Delta(p+k_c)$ & 0 \\
      $S(k_c)$ & 0 & $i\omega_n-\epsilon_{p-k_c}$ & $\Delta(p-k_c)$ & 0 & 0 \\
      0 & 0 & $\Delta(p-k_c)$ & $i\omega_n + \epsilon_{p-k_c}$ & 0 & $-S(k_c)$ \\
      0 & $\Delta(p+k_c)$ & 0 & 0 & $i\omega_n + \epsilon_{p+k_c}$ & $-S(-k_c)$ \\
      $-\Delta(p)$ & 0 & 0 & $-S(k_c)$ & $-S(-k_c)$ & $i\omega_n+\epsilon_p$
    \end{tabular}
  \end{pmatrix}.
\end{equation*}
Here $\omega_n=(2n+1)\pi/\beta$ are the fermionic Matsubara frequencies. The ${\bf X}$ vector contains
the correlation functions defined in the main text, evaluated at same momentum $p$ and frequency $i\omega_n$.
In particular, we have:
\begin{equation}
\mathbf{X}=
\begin{pmatrix}
\begin{tabular}{  c }
$\mathcal G (p,i\omega_n) $ \\
$\mathcal S^\dagger_{+k_c}  (p,i\omega_n) $\\
$\mathcal S^\dagger_{-k_c}  (p,i\omega_n)$ \\
$\mathcal F^\dagger_{\uparrow\uparrow,-k_c} (p,i\omega_n)$\\
$\mathcal F^\dagger_{\uparrow\uparrow,+k_c} (p,i\omega_n)$\\
$\mathcal F^\dagger (p,i\omega_n) $
\end{tabular}
\end{pmatrix}, \qquad
\mathbf{B}=\begin{pmatrix}
\begin{tabular}{  c }
 1  \\
0\\
0\\
0\\
0\\
0\end{tabular}
\end{pmatrix}.
\end{equation}

We now notice that the full solution would require the diagonalization of a large sparse matrix in momentum space.
However, two remarks greatly simplify the equations set. First, the 1D condition of the quantum fluid suggests
to use the nesting property
\begin{equation}
  \varepsilon(k+2k_F)=-\varepsilon(k),
  \label{eq:appnesting}
\end{equation}
where $k_F$ is the Fermi momentum.
This occurs whenever two portions of a Fermi surface are parallel to each other, so that a single wave-vector can connect many points.
The two points $\pm k_F$ are entirely nested at $Q=2k_F$, leading to a diverging
susceptibility in the particle-hole channel, possibly accompanied by a divergence also in the particle-particle
channel, in analogy with Cooper instability~\cite{Giamarchi}. After use of nesting conditions, the system of equations simplifies with:
\begin{equation}
  \mathbf{A}=
  \begin{pmatrix}
    \begin{tabular}{  c c c c c c }
      $i\omega_n-\epsilon_{p}$ & $S(-k_c)$ & $S(\small{k_c})$ & 0 & 0 & $-\Delta(p) $ \\
      $S(-k_c)$ & $i\omega_n+\epsilon_p$ & 0 & 0 & $\Delta(p+k_c)$ & 0 \\
      $S(k_c)$ & 0 & $i\omega_n+\epsilon_p$ & $\Delta(p-k_c)$ & 0 & 0 \\
      0 & 0 & $\Delta(p-k_c)$ & $i\omega_n-\epsilon_p$ & 0 & $-S(k_c)$ \\
      0 & $\Delta(p+k_c)$ & 0 & 0 & $i\omega_n-\epsilon_p$ & $-S(-k_c)$ \\
      $-\Delta(p)$ & 0 & 0 & $-S(k_c)$&$-S(-k_c)$ & $i\omega_n+\epsilon_p$
    \end{tabular}
  \end{pmatrix}.
  \label{eq:AmatrixNest}
\end{equation}
\end{widetext}
A closer look at the system $\mathbf{AX}=\mathbf{B}$ suggests that an algebraic explicit solution cannot be easily extracted,
and that the only remaining possibility would be the diagonalization of a quite large matrix in momentum space.
On the other hand, an explicit ansatz for the superfluid gap function can be introduced as follows,
which then can lead to explicit self-consistent equations for SF and spin order parameters
along with the equation for the chemical potential.

Indeed, one can proceed step-by-step, analyzing the equations for the two order parameters, after setting the other to zero.
What is learned from the solution in the separate regimes is then used as an ansatz for the SF
order parameter in order to derive the full equations. In particular, when setting the spin-wave order parameter to zero,
we notice that the equation for the SF gap function $\Delta(k)$ is
\begin{equation}
\Delta(k)=g\sum_{k'}V_{kk'}\frac{\Delta(k')}{2E_{\rm SF}(k')}\tanh\Big(\frac{\beta E_{\rm SF}(k')}{2}\Big)
\label{eq:bcsgap}
\end{equation}
with the excitation energy
\begin{equation}
E_{\rm SF}(k)\equiv\sqrt{\epsilon_{k}^2+\Delta(k)^2},
\label{eq:excbcs}
\end{equation}
and the number equation
\begin{equation}
N=2\sum_k\Big[\frac{1}{2}-\frac{\epsilon_k}{2E_{\rm SF}(k)}\tanh\Big(\frac{\beta E_{\rm SF}(k)}{2}\Big)\Big]\equiv 2\sum_l n(k).
\label{eq:bcsnumber}
\end{equation}
Equation~\eqref{eq:bcsgap} has solution only when the interaction parameter is positive, $g>0$, since the right-hand-side is a positive-definite quantity.
When the SF order parameter is set to zero instead, the spin-pairing gap equation is:
\begin{equation}
S(Q)=-g\sum_k\frac{S(Q)}{2E_{\rm SDW}(k)}\tanh\Big(\frac{\beta E_{\rm SDW}(k)}{2}\Big),
\label{eq:SQspin}
\end{equation}
where now the excitation energy is
\begin{equation}
E_{\rm SDW}(k)\equiv\sqrt{\epsilon_k^2+\Delta_{\rm SDW}^2},
\label{eq:excspin}
\end{equation}
with $\Delta_{\rm SDW}^2\equiv \sum_{Q=\pm 2k_F} [S(Q)]^2$,
and the number equation is
\begin{equation}
N=2\sum_k \Big[\frac{1}{2}-\frac{\epsilon_k}{2E_{\rm SDW}(k)}\tanh\Big(\frac{\beta E_{\rm SDW}(k)}{2}\Big)\Big].
\label{eq:spinnumber}
\end{equation}
In this case, \eqref{eq:SQspin} has solution only for $g<0$.

Therefore, at least at the present MF level, a second useful remark comes about. By inspection indeed, self-consistency between the left and right members
in the SF gap equation \eqref{eq:bcsgap} requires that $g>0$, while in the SDW equation \eqref{eq:SQspin} requires that  $g<0$. Numerically, the inconsistency shows
up as an oscillation of the right-hand-side member of the two equations at each step of the iterative algorithm, never reaching convergence on the wrong side of the
interaction sign. The opposite signs in front of the $S(q)$ and $\Delta(k)$ terms of $\hat H_{\rm MF}$ would thus imply that no coexistence of SF and SDW phases can
occur at any given $g$ value. This has to be compared with models aimed at explaining the phase diagram of selected high-T$_c$
superconductors~\cite{reviewhtsc, Desta2016}.

We cross-checked this hypothesis by numerically solving the full set of self-consistent equations
with both order parameters in, aided by the following ansatz procedure.
First, given the value of $g$, set $\Delta(k)=0$ and $\Delta_{\rm SDW}$ at the value found from the solution of (\ref{eq:SQspin})-(\ref{eq:spinnumber}) if $g<0$, and set $\Delta_{\rm SDW}=0$ and $\Delta(k)$ at the value found from the solution of (\ref{eq:bcsgap})-(\ref{eq:bcsnumber}) if $g>0$. From the solution of (\ref{eq:bcsgap})-(\ref{eq:bcsnumber}), it is found that the shape of $\Delta(k)$ is characterized by (a) a $4k_F$ periodicity and (b) an overall envelope in the shape of a plateau, exponentially dropping down in size at $|k|> 4k_F$. Assuming that the gross features characterizing the shape of $\Delta(k)$ in $k$ space remain unchanged, an expression for $\Delta(k)$ is then parametrized and inserted in Eq.~(\ref{eq:AmatrixNest}), obtaining an explicit full set of coupled equations for the three unknown quantities.
Inserting the condition $\Delta(k)=\Delta(k+4k_F)$ in an interval $[-2k_F,2k_F]$, and extending the solution to the whole momentum space
with the parametrized envelope function, one obtains the following self-consistent equations:
\begin{widetext}
\begin{eqnarray}
N&=&2\sum_k\Bigg\{\frac{1}{2}-\frac{1}{4}\bigg[1+\frac{\varepsilon_-(k)}{E_s(k)}\bigg] \frac{\epsilon_k}{E_+(k)}\tanh\Big(\frac{\beta E_+(k)}{2}\Big)
-\frac{1}{4}\bigg[1-\frac{\varepsilon_-(k)}{E_s(k)}\bigg] \frac{\epsilon_k}{E_-(k)}\tanh\Big(\frac{\beta E_-(k)}{2}\Big) \Bigg\}, \label{eq:fullcoupledeq0}\\
\Delta(k)&=&\frac{g}{4}\sum_k V_{kk'}\Bigg\{\Bigg[\Delta(k')+\bigg(E_s(k')+\frac{(\Delta(k'))^2-(\Delta(k'-Q))^2}{E_s(k')}\bigg)\Bigg]\frac{\tanh\big({\beta E_+(k)}/{2}\big)}{E_+(k')}\nonumber\\
&&\hspace{2cm} +\Bigg[\Delta(k')-\bigg(E_s(k')+\frac{(\Delta(k'))^2-(\Delta(k'-Q))^2}{E_s(k')}\bigg)\Bigg]\frac{\tanh\big({\beta E_-(k)}/{2}\big)}{E_-(k')}\Bigg\}, \\
S(q)&=&-\frac{g}{4}\sum_k\Bigg\{\frac{S(q)}{E_+(k)}\tanh\Big(\frac{\beta E_+(k)}{2}\Big)+\frac{S(q)}{E_+(k)}\tanh\Big(\frac{\beta E_+(k)}{2}\Big)\Bigg\}.
\label{eq:fullcoupledeq}
\end{eqnarray}
\end{widetext}
In Eqs.~\eqref{eq:fullcoupledeq0}-\eqref{eq:fullcoupledeq}, we have defined the following quantities:
\begin{equation}
  \varepsilon_\pm(k) \equiv \Delta(k) \pm \Delta(k+Q); \quad
  E_s(k)\equiv \sqrt{ \big[ \varepsilon_-(k) \big]^2 + \Delta_{\rm SDW}^2}
\end{equation}
and single-particle energies
\begin{equation}
  E_\pm(k) = \sqrt{\epsilon_k^2+ \big[ \varepsilon_+(k)\pm E_s(k) \big]^2} .
\end{equation}

\begin{figure}[!htp]
  \includegraphics[width=\columnwidth]{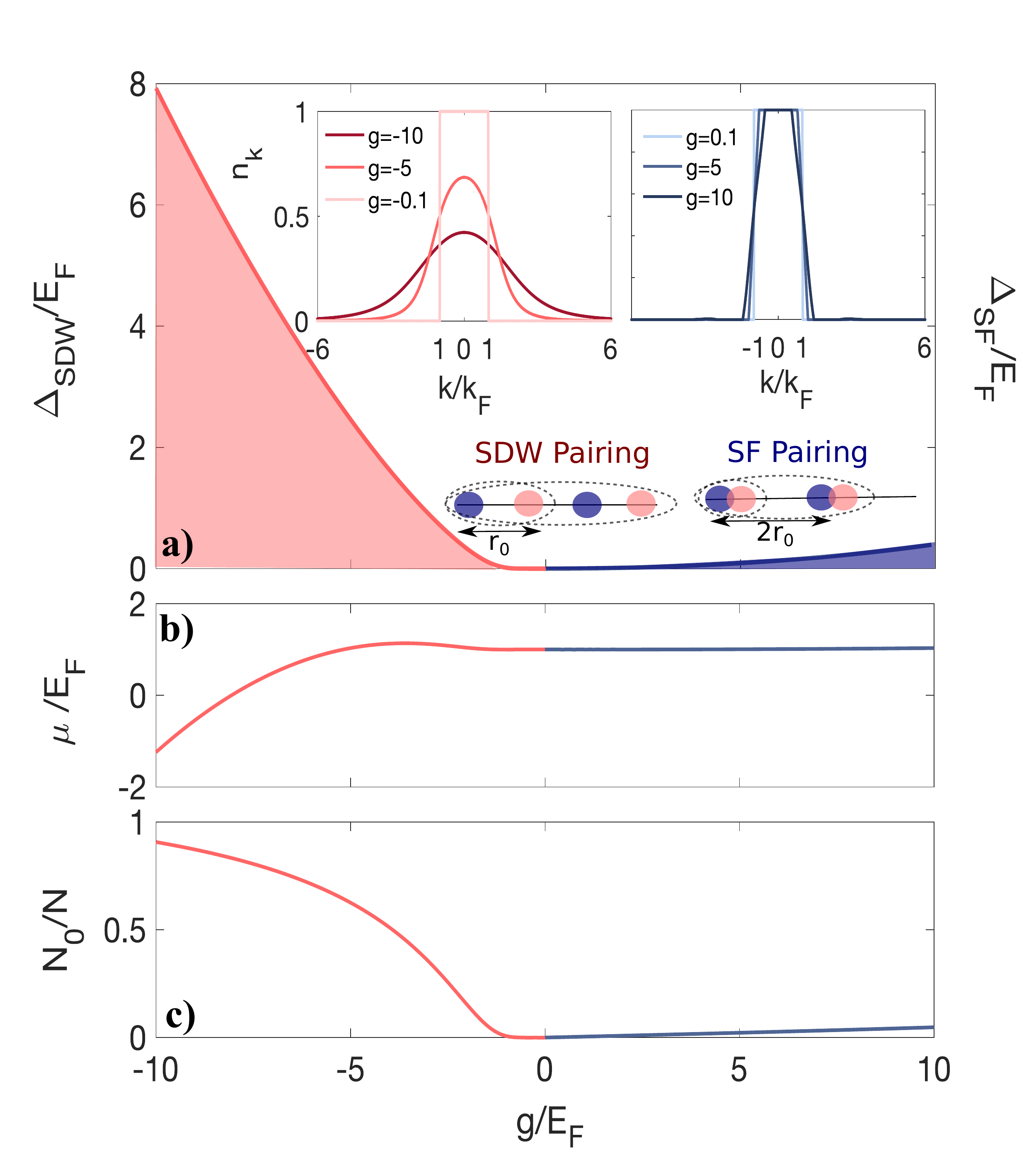}
  \caption{Mean-field phase diagram as a function of $g$.
      \textbf{a}) Order parameters $\Delta_{\rm SDW}$ ($\neq 0$ for $g<0$) and $\Delta_{\rm SF}$ ($\neq 0$ for $g>0$),
      showing no coexistence of SDW and SF phases. Insets: momentum distributions $n(k)$,
      showing bosonic ($g<0$) and fermionic ($g>0$) characters.
      Cartoons represent sketches of fluid structures in terms of average interparticle distance $r_0$,
      as resulting from pair-distribution functions.
      \textbf{b-c}) Chemical potential $\mu$ and corresponding fraction $2N_0/N$ of spin-paired ($g<0$) and SF-paired ($g>0$) particles.}
  \label{fig:fig2}
\end{figure}
The resulting phase diagram in Fig.~\ref{fig:fig2} \textbf{a}) has been obtained by analyzing
$\Delta_{\rm SDW}$ and $\Delta_{\rm SF}$ as functions of the effective coupling $g/E_F$.
Notice the evident asymmetry on the two $g$ sides: this is ascribable to the structure of the original
Hamiltonian, where spin fluctuations represent the main mechanism driving interactions,
directly affecting spin ordering and only indirectly inducing the SF phase.
Figure~\ref{fig:fig2} \textbf{b}) provides further insight, reporting the behavior of the chemical potential $\mu$
sticking to $E_F$ on SF side and dropping to $\mu<0$ at $g\simeq -8$ on SDW side.
The enhancement around $g\simeq -4$ is consistent with mass-renormalization effects.
At the same time, the momentum distributions $n(k)$ (insets), display bosonic-like behavior
on the SDW side accompanied by larger values for the fraction of spin-paired particles in Fig.~\ref{fig:fig2} {\bf c)},
defined as $N_0=\sum_{k,Q} |\langle \hat c^\dagger_{k+Q,\downarrow} \, \hat c_{k,\uparrow}\rangle|^2$ in analogy
with the SF condensate fraction $N_0=\sum_{k}|\langle \hat c_{-k,\downarrow} \, \hat c_{k,\uparrow}\rangle|^2$.
The latter emerges on the SF side with quite smaller values, while $n(k)$ keeps a fermionic character.
Finally, the analysis of pair correlation functions suggests the fluid structures cartooned
in Fig.~\ref{fig:fig2} \textbf{a}). For $g>0$ the SF order parameter evolves from a
periodic structure in $k$-space at weak coupling, arising from the non-local nature of the effective interaction mediated by photons,
towards a standard BCS-like form for stronger coupling.
When $g<0$, a spatially antiferromagnetic-like ordered state develops in the particle-hole channel,
analogous to the companion for $g>0$ in the particle-particle channel: this evolves into a fluid of increasingly
bound composite bosonic pairs of fermions with opposite spins, and eventually a BEC.

\section{Combined exact diagonalization and bosonization analysis}\label{sec:ED}  

In order to explore to which extent large interactions and reduced
dimensionality may drive qualitative changes in the system, we revert back to the microscopic
Hamiltonian~\eqref{eq:H}, and determine the behavior of relevant correlation functions by beyond MF methods.
In fact, we notice that the use of DMRG-based techniques --which would be highly preferable to work out the structure and nature
of the different quantum phases-- is actually unpractical here, because of the presence of long-range interactions.
Therefore we resort to a combination of ED methods through a Lanczos approach, limited to a small number of particles,
and of bosonization analysis, in order to cross-check the results on the fluid structure:
here we are seeking for a cross fertilization of the various approaches.

Exact-diagonalization simulations are performed in momentum space. In order to have comparable numerical data for different values of the number $N$ of fermions,
we tailored the box parameters to make the finite-system density $n$ fixed and coincident with the thermodynamic-limit value $nk_F=2/\pi$. This condition translates
into fixing the mesh $\Delta k= 2\pi/\ell$ to a constant value, where $\ell$ is the system size in real space.
In order to revert to momentum space, we expand
the atomic operators $\hat \Psi^{(\dagger)}_\sigma(x)$ in plane waves:
\begin{equation}
   \hat \Psi_\sigma(x) = \frac{1}{\sqrt{V}} \sum_k e^{i k x} \hat c_{k, \sigma} ,
\end{equation}
where $V$ is the system volume and $c^{(\dagger)}_{k, \sigma}$ (satisfying usual anticommutation relations)
create/destroy a fermion with spin $\sigma$ and momentum $k$.
The effective Hamiltonian~\eqref{eq:H} can thus be written as:
\begin{equation}
   \hat H = \sum_{k, \sigma}\varepsilon_k \hat c^\dagger_{k ,\sigma} \hat c_{k, \sigma} + g \sum_{k, k'} \sum_{Q = \pm k_L}
   \hat c^\dagger_{k+Q, \downarrow} \, \hat c_{k, \uparrow} \, \hat c^\dagger_{k'-Q,\uparrow} \, \hat c_{k', \downarrow}
   \label{eq:Ham_k}
\end{equation}
Notice that, in the MF approximation, the interaction decomposes into a sum of two simpler terms, leading to the breaking of the number
and spin conservation. 

In the discretized space of the momenta and fixed number $N$ of fermions,
each of them may access the two possible spin states $\sigma= \, \uparrow, \downarrow \,$.
We assume that the particles can occupy a discretized set of $L$ different values of the momentum $k$ (sites of the momentum-space lattice).
The Pauli principle forbids to have more than one fermion with a given spin state on the same site, therefore the total Hilbert space size is $4^L$.
Fully exploiting particle number and spin conservation of Eq~\eqref{eq:Ham_k}, we can however restrict the diagonalization to a smaller space.
Specifically, focusing on the zero magnetization sector, our problem reduces to finding the ground state of Eq.~\eqref{eq:Ham_k}
in the effective Hilbert space of dimension ${L \choose N/2}^2$.
In view of the sparseness of the resulting Hamiltonian matrix, we take advantage of a Lanczos-based algorithm, which enables us to
diagonalize matrices of dimensions $\sim (10^7 \times 10^7)$ on a laptop computer without much effort,
as is the case for $N=8, \,L=19$ or $N = 12, \, L = 15$, though for $|g|< 1$ as we have a posteriori checked. We also remark that the maximum momentum
$k_{\rm max}$ that we choose (related to the number of computational sites in $k$-space) mostly depends on the memory size occupied by a single simulation,
which, e.g., for $N=12$ and $L=17$, exceeds 64 Gb. Finally, we remark that simulations performed with both open and periodic boundary conditions show no
appreciable differences.

We computed the ground-state energy and eigenvector, and the density-density
and spin-spin structure factors, i.e., the Fourier transform of correlations functions
$\langle \hat O_\rho(x) \, \hat O_\rho(0)\rangle$ and $\langle \hat O_\sigma(x) \, \hat O_\sigma(0)\rangle
= \sum_{\alpha=x,y} \langle \hat O_s^\alpha(x) \, \hat O_s^\alpha(0) \rangle$,
with the definitions~\cite{Giamarchi}
\begin{eqnarray}
  \hat O_\rho(x) & = & \sum_\sigma \hat \Psi^\dagger_\sigma(x) \, \hat \Psi_\sigma(x), \\
  \hat O^x_s(x) & = & \hat \Psi^\dagger_\uparrow(x) \, \hat \Psi_\downarrow(x) + \hat \Psi^\dagger_\downarrow(x) \, \hat \Psi_\uparrow(x) \\
  \hat O^y_s(x) & = & -i \big[\hat \Psi^\dagger_\uparrow(x) \, \hat \Psi_\downarrow(x) - \hat \Psi^\dagger_\downarrow(x) \hat \Psi_\uparrow(x) \big].
\end{eqnarray}
  
\begin{figure}[!t]
  \includegraphics[width=\columnwidth]{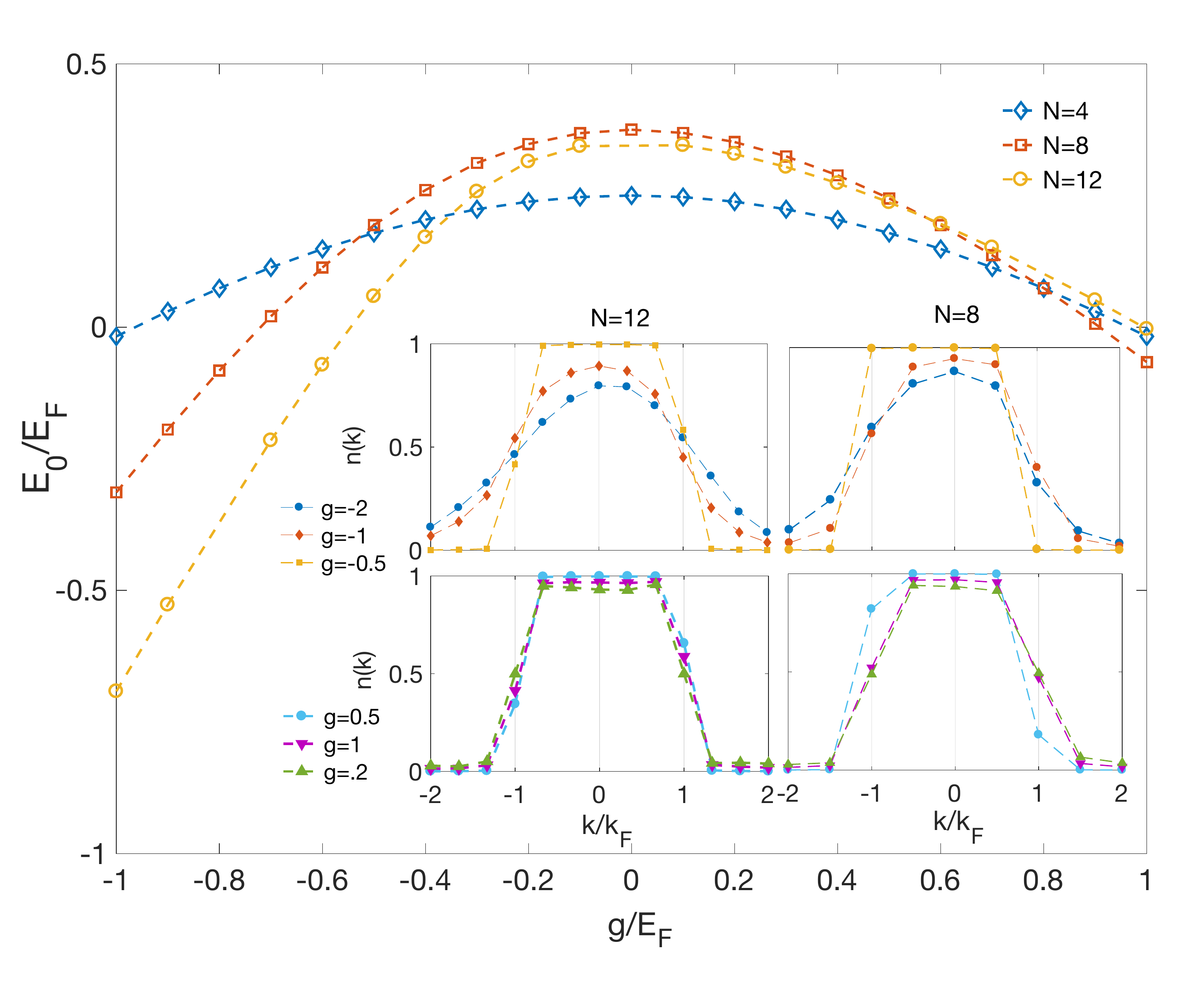}
  \caption{Ground-state energy $E_0$ as a function of the coupling $g$,
      for $N=4, 8, 12$, as obtained with ED,
    showing that the quantum fluid is strongly correlated at all $g$ values.
    Insets: momentum distributions for different $g$ values, confirming the bosonic ($g<0$)
    and fermionic ($g>0$) fluid character of Fig.~\ref{fig:fig2}.}
  \label{fig:fig3}
\end{figure}
\begin{figure}[!t]
  \includegraphics[width=\columnwidth]{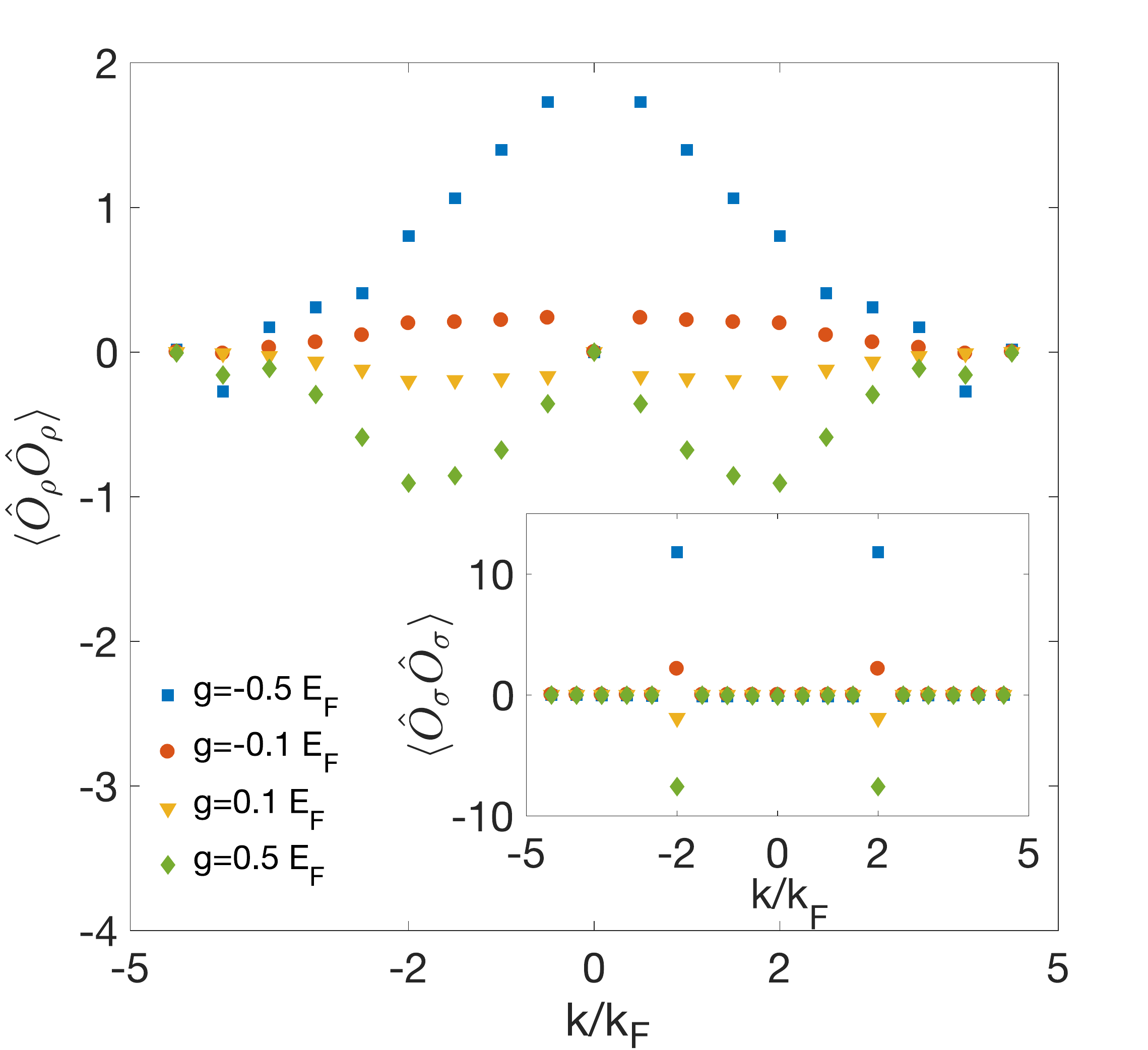}
  \caption{Correlation functions in $k$-space for different $g$ values, from ED with $N=8$.
    In density-density correlations $\langle \hat O_\rho \, \hat O_\rho \rangle$, a peak develops at $k=0$
    for $g<0$ and not for $g>0$, and peaks at $\pm 2k_F$ emerge only for $g>0$, signaling the setting of ADW processes.
    Inset: spin-spin correlations $\langle \hat O_\sigma \,\hat O_\sigma \rangle$; the peaks at $\pm 2k_F$ are
    driven by SDW and ADW processes.}
  \label{fig:fig4}
\end{figure}

Figure~\ref{fig:fig3} displays $E_0$ for different system sizes.
First, one notices the asymmetric behavior, reminiscent of the MF findings,
with the $g<0$ regime energetically favored and the $g>0$ side more rapidly converging to the thermodynamic limit.
Second and most interesting, the quantum fluid is always found in a strongly correlated regime with the ground-state
energy significantly differing from the non-interacting case at all
$g$ values~\cite{Giamarchi}. Finally, the momentum distributions displayed
in the inset, support the bosonic ($g<0$) and fermionic ($g>0$) characters found with MF.
Figure~\ref{fig:fig4} reports density-density $\langle \hat O_\rho \, \hat O_\rho\rangle$
and spin-spin $\langle \hat O_\sigma \, \hat O_\sigma\rangle$ correlation functions.
Notice that $\langle \hat O_\rho \, \hat O_\rho\rangle$ is characterized by a very different behavior
according to the sign of $g$. For $g<0$ a peak structure emerges at $k\sim 0$, which disappears for $g>0$.
The peak is blurred by finite-size effects and develops with increasing $N$, especially for large $g$.
As the $k=0$ value of the density-density correlation is representative of the fluid compressibility,
a diverging value would signal BEC occurrence, consistent with the MF findings.
We assign the peaks developing at $\pm 2k_F$ for $g>0$ to the occurrence of ADW processes.
The larger peaks at $\pm 4k_F$ can be due to Umklapp-scattering processes.
Conversely, $\langle \hat O_\sigma \, \hat O_\sigma\rangle$ displays peaks at $\pm 2k_F$,
expected to be driven by either SDW or ADW processes.
This is confirmed by bosonization analysis, although $k$-space discretization prevents us
to trace their decay law.

The ED analysis has provided definite insights on the bosonic ($g<0$) and fermionic ($g>0$) nature of the quantum fluid across the phase
diagram, and on the primary role played by the spin-density fluctuations in determining its rich physics. On the other hand, the method is limited by the small accessible
values of $N$ and - at the same time - the need of working in a canonical ensemble, i.e. with fixed number of particles, where the SF phase-definite quantum
ground-state cannot be built up. To overcome these limitations, we combine ED with a bosonization analysis of the low-energy physics of
Hamiltonian~\eqref{eq:H}. To this aim, we introduce
the bosonized expression of the fields for the fermionic operators~\cite{Giamarchi},
$\hat \Psi_\sigma(x) = e^{-ik_Fx} \hat \Psi_{R\sigma}(x) + e^{ik_Fx} \hat \Psi_{L\sigma}$,
where $\hat \Psi_{R(L)\sigma}(x)$ are the right and left movers,
respectively, i.e.
\begin{align}
  \hat \Psi_{r,\sigma}(x) = \frac{1}{\sqrt{2\pi\alpha}} &
  \exp \Big( \frac{i}{\sqrt{2}} \Big[ \hat \theta_\rho(x)-r \hat \phi_\rho(x) \nonumber \\
    & + \sigma \big( \hat \theta_\sigma(x) - r \hat \phi_\sigma(x) \big) \Big] \Big) ,
\end{align}
where  $r=\pm$ standing for right/left movers, $\sigma = \, \uparrow, \downarrow$, and $\alpha$ is the cutoff.
Here, $(\hat \phi_\rho, \hat \theta_\rho)$ and $(\hat \phi_\sigma, \hat \theta_\sigma)$ represent
the bosonic fields that are linked to the charge and spin channel respectively.
The resulting bosonized expression of Hamiltonian (1) for $k_L=2k_F$ is
\begin{eqnarray}
  \hat H_{B} & \sim & \sum_{l=\rho,\sigma}\frac{1}{2\pi}\int dx \Big( u_lK_l \big[ \pi \hat \Pi_l(x) \big]^2+\frac{u_l}{K_l} \big[ \nabla \hat \phi_l(x) \big]^2 \Big) \nonumber \\
  & & + \frac{g}{4\pi^2\alpha^2} \int \!\! \int dx \, dx'
  e^{-i\sqrt{2}(\hat \theta_\sigma(x) - \hat \theta_\sigma(x'))} \nonumber \\
  && \hspace{2cm} \times \cos \big[\sqrt{2}(\hat \phi_\rho(x) - \hat \phi_\rho(x') ) \big] ,
\end{eqnarray}
with $\hat\Pi_l=\nabla\hat\theta_l$. Here, $K_\sigma$ and $K_\rho$ are the Luttinger parameters for the corresponding two channels.
In the present case, $u_l=v_F$ is the Fermi velocity and $K_l=1$ corresponds to the noninteracting case.
In the absence of an optical lattice, all terms involving a highly oscillating behavior average to zero,
so that only slowly varying terms are retained.
The main feature of the interaction Hamiltonian is its non-locality, embedded in the long-range nature of photon mediated
interaction, which involve not only the spin sector via the $\hat \theta_\sigma$ field but also the charge sector,
through the $\hat \phi_\rho$ field.

To get a qualitative picture of the low-energy behavior we use a MF type
of argument, analyzing separately the case in which either the field $\hat \theta_\sigma$ or $\hat \phi_\rho$ becomes massive.
When the spin becomes massive, opening a gap in the excitation spectrum $\Delta= \langle e^{i \hat \theta_\sigma}\rangle$,
the system tends to open a gap in the charge sector when the operator $\cos (\sqrt{2} \hat \phi_\rho)$ becomes relevant,
i.e. $K_\rho <3$ and this behavior is independent on the sign of the interaction.
On the other hand, when $\hat \phi_\rho$ becomes massive and is trapped in one of the minima
of the cosine $\Gamma=\langle \cos(\sqrt 2 \hat \phi_\rho)\rangle$ a gap in the spin sector is opened when $K_\sigma>1/3$
which makes the term $\cos (\sqrt{2} \hat \theta_\sigma)$ relevant for $g>0$.
Thus for $g>0$ the instabilities of the system can be both ADW and SDW.

In particular, we look at the following correlation functions, characterizing ${\rm SDW}_{x-y}$, ADW and singlet (SS)/triplet (TS) superfluidity~\cite{Giamarchi}:
$\big\langle \hat O_{\rm SDW}^{x-y \dagger}(r) \, \hat O_{\rm SDW}^{x-y}(0) \big\rangle$, 
$\big\langle \hat O_{\rm ADW}^\dagger(r) \, \hat O_{\rm ADW}(0) \big\rangle$,
$\big\langle\hat O_{\rm SS}^\dagger(r) \, \hat O_{\rm SS}(0) \big\rangle$, and
$\big\langle \hat O_{\rm TS}^\dagger(r) \, \hat O_{\rm TS}(0) \big\rangle$.
Here, the operators are defined as
\begin{eqnarray}
\hat O_{\rm SDW}^{x} & = & \hat \Psi_{R\uparrow}^\dagger \hat\Psi_{L\downarrow}+\hat \Psi_{R\downarrow}^\dagger \hat\Psi_{L\uparrow}\\
\hat O_{\rm SDW}^{y} & = & -i(\hat \Psi_{R\uparrow}^\dagger \hat\Psi_{L\downarrow}-\hat \Psi_{R\downarrow}^\dagger \hat\Psi_{L\uparrow})\\
\hat O_{\rm ADW} & = & \hat \Psi_{R\uparrow}^\dagger \hat\Psi_{L\uparrow}+\hat \Psi_{R\downarrow}^\dagger \hat\Psi_{L\downarrow}\\
\hat O_{\rm SS} & = & \hat \Psi_{R\uparrow}^\dagger \hat\Psi_{L\downarrow}^\dagger+\hat \Psi_{L\uparrow}^\dagger \hat\Psi_{R\downarrow}^\dagger\\
\hat O_{\rm TS}&=&\hat \Psi_{R\uparrow}^\dagger \hat\Psi_{L\downarrow}^\dagger-\hat \Psi_{L\uparrow}^\dagger \hat\Psi_{R\downarrow}^\dagger,
\end{eqnarray}
We then look at the Fourier transform of the above correlation functions, that is
\begin{equation}
\chi_a (k,\omega_n) =\! \int \! dr \int_0^\beta \! d\tau \,e^{(-ikr+i\omega_n \tau)} \big\langle \hat O_{a}^\dagger(r, \tau) \, \hat O_{a}(0) \big\rangle
\end{equation}
and, after performing analytical continuation, we end up with the susceptibilities.
From the behavior of these susceptibilities we  can gain insight on the nature of the quantum fluid and on its phase diagram. In particular, when the cavity field is
red-detuned with respect to the pump frequency ($g<0$), from the RG analysis we find values of $K_\rho$ and $K_\sigma$ that lead to divergent behavior
in the ${\rm SDW}_{x-y}$ correlation function, which for $k=2k_F$ it diverges as
$\chi_{{\rm SDW}_{x-y}}\sim \omega^{K_\rho+K_\sigma^{-1}-2}$. On the blue-detuned side ($g>0$) instead, the main instabilities are in the SS channel, with
$\chi_{\rm SS}\sim \chi_{{\rm TS}}\sim \omega^{K_\rho^{-1}+K_\sigma-2}$. No sign of ADW is found at this level,
so processes observed in the ED results can be explained as higher-energy processes which cannot be captured at the
lowest order of our bosonization of the fields operators~\cite{LongPaper}.

\section{Concluding remarks}\label{sec:conclusions}
We showed that degenerate fermionic atoms in optical cavities may stabilize a spinful Fermi fluid
in strongly-coupled conditions, for both positive and negative values
of the experimentally tunable effective interaction $g$ mediated by cavity-photons.
Different quantum phases, driven by the most favored processes (spin-density fluctuations) appearing in the original Hamiltonian, can be explored.
These in turn, appear to indirectly favor the occurrence of fluctuations
in the atomic density and superfluid pairing, leading to a rich phase diagram
where a SDW structure with bosonic nature emerges for $g<0$, while a superfluid character (singlet or triplet)
with fermionic nature is favored for $g>0$.
Our findings synthesize different clues extracted from mean-field and exact-diagonalization,
accompanied by bosonization and renormalization-group analysis.

Our proposed system can be implemented in current experiments using ultracold fermionic atoms in optical cavities,
under conditions of large detuning and bad-cavity regimes~\cite{EsslingerReview}, and confined 1D geometry realized by
a tight confinement along the directions perpendicular to the cavity axis with a superimposed trapping~\cite{EsslingerOpto}.
Typical values include $\kappa\approx$ MHz, $E_R=\hbar^2 k_L^2/(2m)\approx$ kHz, comparable with the $E_F$
of $N \sim 10^5$ atoms, $\Delta_c \sim 10-30$ MHz. With these numbers, the two-photon cavity frequency $g_{\rm eff}$ can be tuned in strength
by changing the intensity of the transversal pump and then vary the sign and strength of $g$ acting on the cavity detuning.

Further developments are needed to characterize the nature and dominance of the different order parameters
across the phase diagram, as using twisted boundary conditions to mimic superfluid response
not accessible in the canonical ensemble, or more sophisticated methods to increase the system size
in presence of long-range interactions. Also, optical cavities are by no means closed systems,
requiring to investigate the phase-diagram stability towards dissipative effects.
The system studied here is in fact equivalent to a one-axis twisting Hamiltonian
that can be used to engineer spin-squeezing mechanisms~\cite{SquezRev}.

\acknowledgments
We thank M. Holland, T. Esslinger, and V. Vuletic for useful discussions.
We thank H. Ritsch also to have pointed out the possible use of a ring cavity~\cite{refnoteHelmut}.
This work has been realized within the MAGIA-Advanced project (INFN) and the MIT-UniPi Program.

\bibliographystyle{apsrev}

\begin{thebibliography}{33}
\bibitem[{ref({\natexlab{a}})}]{refnoteHelmut}
\bibinfo{note}{Starting from a ring cavity, the same
  Hamiltonian $\hat H_{\rm MF}$ can be obtained without further approximations.}
\end{thebibliography}

\end{document}